# Title: Antiskyrmionic ferroelectric medium

**Authors:** Mauro A. P. Gonçalves[1], Marek Paściak[1], Jiří Hlinka[1]*

**Affiliations:**
[1]Institute of Physics, Czech Academy of Sciences; Prague, 182 21, Czech Republic.

*Corresponding author. Email: hlinka@fzu.cz

**Abstract:**

Typical magnetic skyrmion is a string of inverted magnetization within a ferromagnet, protected by a sleeve of a vortex-like spin texture, such that its cross-section carries a positive integer topological charge. Some magnets form antiskyrmions, the antiparticle strings which carry a negative topological charge instead. Here we demonstrate that topologically equivalent but purely electric antiskyrmion can exist in a ferroelectric material as well. In particular, our computer experiments reveal that the archetype ferroelectric, barium titanate, can host antiskyrmions. The polarization pattern around their cores reminds ring windings of decorative knots rather than the typical magnetic antiskyrmion texture. We show that the antiskyrmion of barium titanate has just 2-3 nm in diameter, a hexagonal cross-section, and an exotic topological charge of minus two. We deduce that formation of antiskyrmions is favored by a fortunate combination of the moderate anisotropy of the anharmonic electric susceptibility and the characteristic anisotropy of the polarization correlations in barium titanate crystals.

**Significance statement:**

The prediction and experimental confirmation of magnetic skyrmions - the objects with nontrivial swirling spins patterns - revolutionized the physics of nanoscale magnetism and opened new horizons for spintronics. In spite of the inherently shorter and faster correlations of the electric polarization, the recent developments in electric skyrmionics follow these innovations. This work reveals that classical ferroelectric perovskite - $BaTiO_3$ - can host 2-3 nm wide polar columns spontaneously surrounded by a unique noncollinear polarization pattern that has never been described before. We analyze how this pattern is formed and stabilized. Based on its negative integer topological charge, we name this soliton as "ferroelectric antiskyrmion".

**One-Sentence Summary:**

Ultrasmall ferroelectric domains in barium titanate carry a negative topological charge.

**Main Text:**

Discovery of the topologically stable nanoscale swirling defects of magnetization textures in chiral ferromagnets [1,2], termed as magnetic skyrmions, has opened amazing perspectives for information storage technology [3] as well as a vast arena for studies of intriguing new physics phenomena [4,5] like the topological Hall effect [6] and the skyrmion Hall effect [7–9]. It is now well understood that skyrmions and similar topological defects can be stabilized in many other, sometimes even achiral or centrosymmetric, magnets and antiferromagnets [10–15].

Geometrically analogous nanoscale swirling defects of electric polarization textures have been also observed in nonmagnetic ferroelectric materials [16]. They have been attracting attention due to their truly nanoscale size [16–18] and intriguing physical properties such as negative capacitance [19] or emergent chirality [20–24]. Experimental evidence of ferroelectric skyrmions in nanostructures consisting of nanoscale layers of ferroelectric $PbTiO_3$ combined with paraelectric layers of $SrTiO_3$ [16] pointed towards the basic role of the interface effects. These observations then raised a question about whether such ferroelectric skyrmion can be stable in the bulk of the ferroelectric material itself.

Natural candidate materials are those in which the electric polarization is expected to rotate gradually within the thickness of their domain walls. Fig. 1a shows a plausible geometry of a columnar nanodomain in bulk $PbTiO_3$. The nanodomain, defined by the polarization parallel to $z$ direction, is surrounded by a matrix of the opposite polarization. Computational study of Ref. [17] suggests that in $PbTiO_3$, an in-plane polarization develops at the domain wall, forming a closed loop. The resulting polarization arrangement corresponds to a stable Bloch-like skyrmion. Our original motivation was to verify whether a similar nanodomain can persist in the rhombohedral phase of $BaTiO_3$. The antiskyrmion reported here is such a stable nanodomain, but the geometry and topological charge of its polarization pattern is actually very different from that of $PbTiO_3$. We are showing here that the peculiarity of this $BaTiO_3$ nanodomain consists in its antiskyrmion nature.

**Topological and crystallographic considerations**

Both magnetic and ferroelectric skyrmions are string-like defects in a vector order parameter field. The non-trivial topology of the skyrmion is characterized by a non-zero skyrmion number $N_{sk}$ given as an integral [18,25,26] of the topological density $q(x, y) = (1/4\pi)\, \mathbf{u}\cdot(\partial_x \mathbf{u} \times \partial_y \mathbf{u})$ over the $xy$ plane perpendicular to the skyrmion string, where $\mathbf{u}$ is a unit vector field defining the direction of the order parameter. Topological charge with an invariant sign can be defined as $Q = u_z(0)\cdot N_{sk} = - u_z(\infty)\cdot N_{sk}$, where $u_z(0)$ and $u_z(\infty)$ are the $z$-components of the field direction vector in the core of the defect and far away from it. This quantity $Q$ is positive for skyrmions and negative for antiskyrmions. In case of ferroelectric skyrmions the vector field $\mathbf{u}$ is typically obtained from the local electric dipole of each unit cell and the discrete version of topological charge expression is used instead of the integration [17,18]. In particular, the $PbTiO_3$ skyrmion shown in Fig. 1a yields its invariant topological charge of $+1$, common to Néel magnetic skyrmions in $GaV_4S_8$ [27] or Bloch magnetic skyrmions in MnSi [1,4]. In contrast, we report in Fig. 2 a stable ferroelectric nanodomain of $BaTiO_3$, carrying a topological charge of $Q = -2$.

Plausible overall geometry of nanodomains of antiparallel polarization is sketched in Fig. 1. It follows from electrostatic considerations that domain boundaries should be parallel to the direction of the spontaneous ferroelectric polarization. The cross-section of the nanodomain should agree with the macroscopic crystal symmetry of the ferroelectric domain itself and in case of strong anisotropy of the domain wall energy it would be a polygon aligned with the crystallographic axes.

To give a more precise description of our trial and relaxed configurations, we introduced ($x$, $y$, $z$) Cartesian systems for PbTiO$_3$, while ($x'$, $y'$, $z'$) refers to BaTiO$_3$. Crystallographic directions and planes are described with Muller indices $h$, $k$, $l$ of the parent cubic structure. We note that for the tetragonal ferroelectric PbTiO$_3$, $x \parallel [100]$ and the unique ferroelectric axis is $z \parallel [001]$, while in the rhombohedral groundstate of BaTiO$_3$, $x' \parallel [01\text{-}1]$, and the unique ferroelectric axis is set as $z' \parallel [111]$. The brackets with an asterisk superscript ({ }*) are used to denote the subset of planes equivalent within the symmetry of the given *ferroelectric domain* state (say, that of the core). For example, the {100}* set for PbTiO$_3$ domain with polarization along $z$ contains only (100) and (010) planes. These low-index planes are those of PbTiO$_3$ domain walls shown in Fig. 1a. Similarly, the stable nanodomains in BaTiO$_3$ should be delimited either by the set of {1-10}* facets or by the set of {2-1-1}* facets. Note that both sets contain planes parallel to the ferroelectric axis $z'$. Using the convention of Ref. [28], we can alternatively state that PbTiO$_3$ skyrmions are delimited by the neutral T180{100} domain walls (that is, the {100}*-oriented walls in *T*etragonal phase corresponding to *180*-degree rotation of the spontaneous polarization), while the stable nanodomains in rhombohedral BaTiO$_3$ are expected to be delimited by either R180{1-10} domain walls (as in Fig. 1b) or by the R180{2-1-1} domain walls (as in Fig. 1c).

It is worth noting that a necessary requirement for nonzero $Q$ is a non-coplanar order parameter texture. Clearly, the non-coplanarity of the polarization profile of the PbTiO$_3$ nanodomain in Fig. 1a is related to the anticipated non-colinear, Bloch-like (chiral) structure of its T180{100} domain walls [29]. Likewise, in BaTiO$_3$, according to the phenomenological Ginzburg-Landau theory and DFT calculations, both R180{1-10} and R180{2-1-1} domain walls should have a strongly non-colinear character [28,30–32]. In this respect, rhombohedral BaTiO$_3$ appeared to be a natural candidate for searching interesting topological defects.

**Computer experiment**

Since the exact shape and respective energy of antiparallel domain walls of rhombohedral BaTiO$_3$ happens to be sensitive to the parametrization of the Ginzburg-Landau potential [31,32], we have employed here a more reliable *ab initio*-based atomistic shell model potential optimized for BaTiO$_3$ molecular dynamics simulations [33,34]. This model has been successfully applied to reproduce various experimental data, including anomalous thermal diffuse scattering patterns in its cubic phase [35] or short-range correlations in amorphous BaTiO$_3$ [36]. Each atom is represented by core and shell, with its own position and electric charge. The potential energy includes pairwise short-range interactions as well as electrostatic Coulomb interactions. We have used supercells made of 12×12×4 hexagonal unit cells (*i.e.*, 8640 independent atoms) with periodic boundary conditions. Stable topological defects in the ferroelectric ground state were searched for through relaxing various initial configurations using classical molecular dynamics simulations at the temperature of 1 K.

In order to avoid any bias towards the nonzero topological charge, most of our initial configurations were colinear polarization textures. We have started with a homogeneously polarized supercell with $P_{z'} < 0$ and inverted polarization in a columnar region of hexagonal cross-section, corresponding to Fig. 1b. The inner diameter $D$ of the hexagonal area of positive $P_{z'}$ is given by $D \approx N.d$, where $d$ is the nominal spacing of the (1 -1 0) Ti planes in the parent cubic phase, and $N$ is the dimensionless diameter of the hexagon.

**Resulting polarization texture**

The relaxed structure, obtained from the initially topologically trivial ($Q = 0$) bubble nanodomain with $N = 11$ and $D \approx 3$ nm, is shown in Fig. 2a. While the overall columnar shape persisted and the $P_{z'}$ components did not change much, considerable in-plane polarization components developed spontaneously around the circumference of the nanodomain. Let us stress that the magnitude of the in-plane polarization is comparable to the value of the single-domain spontaneous polarization (see the labeling in Fig. 2a).

We note that also the radial and tangential *gradients* of the polarization are comparable in magnitudes. Therefore, the above anticipated domain wall picture (Fig. 1b and c) is actually not much adequate: the domain wall width would be about the same as the length of the individual wall segment. Rather, the in-plane polarization texture, represented by the arrows in Fig. 2a, is a condensate of six different vortices, with alternatively clockwise and anticlockwise sense of polarization rotation, each centered at one of the edges of the $P_{z'} > 0$ hexagon. Fig. 2b shows the simplified sketch of the polarization arrangement obtained, emphasizing the interlinked components of the reversed out-of-plane and swirling in-plane polarization.

Fig. 2c shows the topological density of this relaxed state, which presents six pronounced local minima in the center of each vortex and six rather small maxima near the vertices of the hexagon. Contributions of the individual equilateral triangular plaquettes in the hexagonal lattice of the projected Ti ion positions correspond to the exact values of the oriented solid angle subtended by the three dipoles in the corners of these elementary plaquettes. We have checked that none of the dipoles is zero and none of the plaquettes yields the corresponding solid angle critically close to the borderline value of $2\pi$. Therefore, summing up all these individual contributions, the total topological charge of $Q = -2$ was obtained without any ambiguity. This is why this topological defect deserves to be called ferroelectric antiskyrmion.

**Stability of the ferroelectric antiskyrmion**

The relaxed configuration shown in Fig. 2 corresponds to the local energy minimum – its stability with respect to infinitesimal perturbations has been confirmed by checking the sign of all eigenvalues of the dynamical matrix corresponding to the internal lattice degrees of freedom of the supercell. To further appreciate the robustness of the antiskyrmion configuration, we have also attempted to erase the nanodomain by a strong electric field aligned along the $z'$ axis. It turned out that the nanodomain is stable in a very broad range of $-850$ kV/cm $< E'_z < 1000$ kV/cm. In this range, the topological charge is always equal to -2. Since the critical fields are only 4-5 times less than the ideal coercive field limit for the homogeneous switching of the single domain state ($E_c = 4600$ kV/cm), we conclude that this ferroelectric antiskyrmion is a remarkably resistant object.

**Uniqueness of the antiskyrmion geometry**

We have noticed that the diameter of the antiskyrmion defining the width of the $P_{z'} > 0$ area is comparable with the range of the smearing of the $P_{z'}(x')$ domain wall profiles (Fig. 2d). Therefore, we have attempted to stabilize antiskyrmions of different sizes by varying the size of the initial topologically trivial nanodomains (see Fig. 3). The smallest antiskyrmion that we could create had $N = 9$ (see Fig. 3b). For smaller initial diameters, the nanodomain decays back to the single domain ground state (see Fig. 3a). The $N = 13$ antiskyrmion is shown in Fig. 3. All solutions share the $Q = -2$ topological charge and similar geometry.

We have also explored other initial nanodomain configurations. In general, we have observed that simple inversion of $P_{z'}$ within a compact columnar nanodomain with a diameter of 3-4nm typically trigger a natural evolution towards the $Q = -2$ state. Despite certain structural variations, the persistent topological defects share the same overall geometry (see Fig. 4c, 4i, 4l and 2a). In particular, the final antiskyrmion texture does not depend on whether the initial nanodomain has R180{1 -1 0} or R180{2 -1 -1} domain walls (compare Fig. 2a and 4a). In other cases, the initial nanodomains evolved to the single domain state. For instance, the configuration similar to Fig. 2a but with inverted in-plane polarization is not stable (see Fig. 4d-f, note that flipping $y'$ to $-y'$ is not a symmetry operation of the parent phase). In overall, these results indicate that the antiskyrmion solution is much more stable than any other columnar nanodomain that we have tested.

**Discussion**

It is worth noting that magnetic antiskyrmionic defects with $Q < 0$ are known [39], and they can even be derived from the canonical Bogdanov-Yablonskii phenomenological theory [40,41]. Nevertheless, the parent paraelectric phase of $BaTiO_3$ is centrosymmetric, so that the formal ferroelectric analogue [42] of the bulk Dzialoshinskii-Moriya interaction is forbidden. Therefore, the Bogdanov-Yablonskii theory [40] does not explain the existence of the antiskyrmions investigated in this work, and we can also exclude the interfacial Dzialoshinskii-Moriya mechanism [43] since there is no interface in our simulations.

Likewise, the bulk skyrmion stabilization scenario of $PbTiO_3$, based on the emergence of Bloch character of the domain walls, does not apply to the formation of antiskyrmions in the rhombohedral ferroelectric phase of $BaTiO_3$, because the observed antiskyrmion is not composed of Bloch walls. In fact, the polarization profile of the "Bloch" (tangential in-plane) polarization component changing its sign in the vicinity of the hypothetical domain wall center (see $P_{y'}(x')$ in Fig. 2d), rather than reaching its maximum absolute value like in the $PbTiO_3$ case.

This so called bichiral domain wall profile [44,45] has been previously found to have an energy very close to that of Bloch solutions in planar domain walls of $BaTiO_3$ [28,30], which further corroborates the appearance of this feature within the antiskyrmion. Generalizing the conclusions of a more rigorous analysis made for selected planar bichiral and Bloch domain walls [46], we can infer that stability of the R180{1-10} bichiral domain walls is favored by a moderate anisotropy of the anharmonicity of the parent phase electric susceptibility (favoring polarization rotation) combined with a pronounced anisotropy of the polarization correlations. Similarity of bichiral walls and the $P_{y'}(x')$ profile of Fig. 2d allows us to infer that also the observed antiskyrmion pattern owes its existence to the combination of the moderate anisotropy of the anharmonic electric susceptibility of $BaTiO_3$, related to the competition of its tetragonal, orthorhombic and rhombohedral ferroelectric phases, and the pronounced anisotropy of polarization correlations of $BaTiO_3$, manifested also in its highly anisotropic soft phonon dispersion relations [47] and in its peculiar diffuse scattering pattern [35].

Finally, the most interesting is the polarization patterns in the corners of the $P_{z'}>0$ hexagon, where the most pronounced in-plane polarization is found, and where it has a dominantly radial orientation. Unlike in the Néel skyrmion, however, this radial component is alternatively pointing outwards and inwards. The sense of this radial (Néel) polarization is fixed by the underlying parent cubic crystal anisotropy. The essential role of this anisotropy on the stability of the antiskyrmion was demonstrated by the collapse of the flipped pattern shown in Fig. 4d. It is this moderate but essential anisotropy what also dictates the the sense of rotation and the number of turns of the in-plane polarization encountered when going around the circumference of the $P_{z'}>0$ hexagon, what eventually results in $|Q| > 1$.

While we are not aware of any investigations of a similar ferroelectric antiskyrmion defects, the higher-skyrmion number $|Q| > 1$ have been already considered for magnetic systems [48–52]. The geometrically closest textures described in the literature are probably the $Q = -2$ motifs in the magnetic textures of hexagonal (also called triangular) magnets [49,50] and ultrathin magnetic films [53,54]. Although the magnetic interactions are very different from the ferroelectric ones, phases of hexagonal magnets may serve as an inspiration for investigations of the condensation of the isolated ferroelectric antiskyrmions into the ferroelectric antiskyrmionic lattices. One of the remarkable common aspects of these magnetic and ferroelectric antiskyrmions is that the topological density is concentrated only in a few "hot spots" with fractional topological charge [18,49,55]. In the present ferroelectric antiskyrmion, there are six such hot spots, each carrying a large part of the "quark-like" $Q \approx -1/3$ topological charge of the corresponding vortex segment of the entire antiskyrmion.

**Conclusion**

In summary, we studied extremely small columnar ferroelectric nanodomains in the low-temperature ferroelectric phase of BaTiO$_3$ by means of established ab-initio based atomistic methodology [33–34,37–38].

We realized not only that the ferroelectric nanodomains with a miniscule 2-3 nm diameter can be stabilized, but also that these nanodomains have a characteristic hexagonal shape and that six vortices develop, each on one facet of the nanodomain, and each carrying a fractional topological charge of about -1/3, giving altogether the unusual net skyrmion number of -2. We computationally affirmed the stability of these ferroelectric antiskyrmions, their remarkable resilience to the electric field and explored the kinetics of their formation.

This work indicates that nanoscale ferroelectric topological defects exists not only in PbTiO$_3$, but also in other ferroelectric materials, and that distinct ferroelectric materials can carry completely different kinds of truly nanoscale topological defects. We believe that these findings will propel further experimental and theoretical developments of ferroelectric skyrmionics.

**Methods**

*Interatomic potential.* For all atomistic calculations/simulations in this study we used the interatomic potential originally developed for atomistic molecular dynamics solid solutions of of BaTiO$_3$ and SrTiO$_3$ [33]. It has been parametrized considering mostly energies, forces, stresses, and phonon properties of a set of properly chosen auxiliary atomic configurations, calculated within the density functional theory [34]. Parameters of the model adopted here are those taken from Ref. [34]. This potential utilizes the framework of the shell model, which by representing each atom with a core and a shell mimics atomic polarizability, and with the anharmonic core-shell interaction is particularly well-suited to reproduce the ferroelectric behavior [33]. In particular, the parametrization predicts correct sequence of BaTiO$_3$ phase transitions (with the cubic-tetragonal phase transition temperature at $T_C \approx 360$ K and the rhombohedral ferroelectric phase explored here stable up to $\sim 140$ K [33]).

*Nanodomain construction and relaxation.* We worked with supercells made of 12×12×4 rhombohedral BaTiO$_3$ unit cells in the hexagonal setting (*i.e.* 8640 atoms) subjected to periodic boundary conditions. The initial polarization configurations were set by displacing the Ti atoms from the centers of oxygen octahedra along the $z'$ direction (positive displacement in the nanodomain and negative in the matrix). The Ti atoms in the one-atom-thick layer at the domain boundary remained in the ideal cubic positions.

Classical molecular dynamics (MD) simulations were taken as the main engine for structure relaxation of the studied systems. To this end the DL POLY software [37] was used and the calculations were done with constant strain – constant temperature ensemble, 0.4 ps timestep and 10 ps initial equilibration time. Temperature was set to 1 K to allow the structures for effective relaxation involving small amount of thermal noise. The systems were kept evolving for several tens of ps, until no change in the time-averaged local dipole configurations could be observed (usually after ∼10 ps of the production run).

To confirm that the obtained relaxed configurations are in energetic minima, the atomic positions were subjected to rational function optimization method as implemented in the program GULP [38]. The method involves diagonalization of the Hessian and it follows the modes with the smallest eigenvalues in the optimization process. We found that the MD-derived structures needed no meaningful steps in the optimization with the minuscule reduction of the energy (<0.1 eV per 5-atom formula unit). Final check of stability comprised calculation of a dynamical matrix and its eigenvalues (also within GULP) to exclude any remnant unstable modes.

Local dipole moments of structures chosen for the presentation were evaluated for each 5-atom perovskite cell separately on the basis of positions and charges of cores and shells. For a given cell one Ti atom with its surrounding octahedron of six O atoms (with the weight 0.5) and 8 coordinating Ba atoms (0.125 weight) was considered.

**Acknowledgments:**

**Funding:**

This work was supported by the Czech Science Foundation (project no. 19-28594X).

**Author contributions:**

JH proposed the scope of the investigations and the outline of the paper; MAPG performed the ensemble of simulations under the guidance of MP; all authors discussed the intermediate progress and all equally contributed to the final manuscript.

**Competing interests:**

Authors declare that they have no competing interests.

**Data and materials availability:**

All data are available in the main text or upon a reasonable request from authors

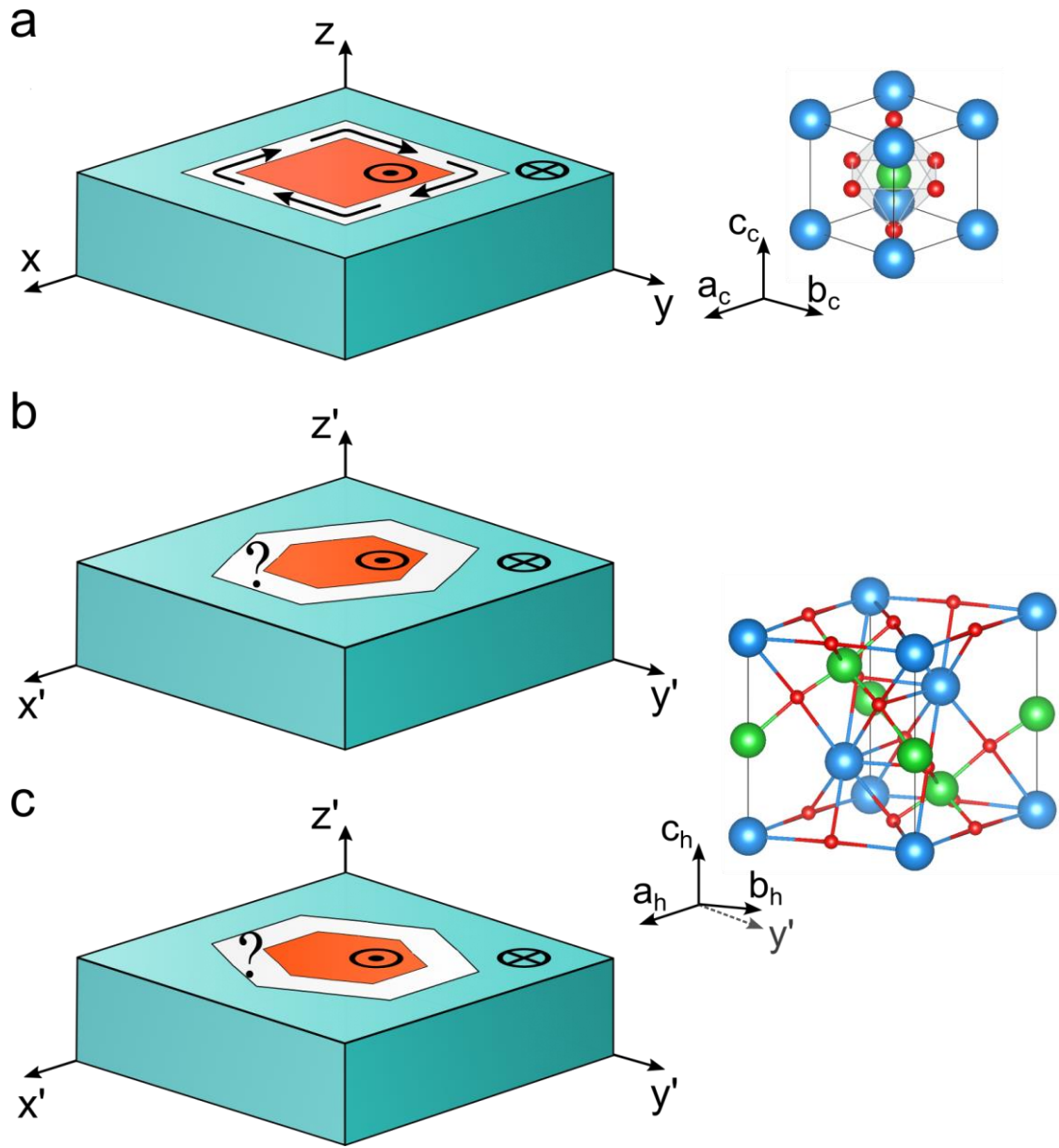

**Fig. 1. Ultrasmall domains of antiparallel polarization in PbTiO$_3$ and BaTiO$_3$.** (a) Typical nanodomain configuration in tetragonal PbTiO$_3$ as identified in previous investigations [17]. The polarization of the inner part (orange) is parallel to the ferroelectric $z$ axis, the outer one (turquoise) has the opposite polarization. Black arrows in the domain wall region (white) represent the inner domain wall polarization forming a close loop around the nanodomain. (b) The anticipated shape of nanodomain within a rhombohedral BaTiO$_3$ with the ferroelectric axis $z'$ ∥ [111]. Domain walls are {1-10}* crystallographic planes, one of them is perpendicular to $x'$ axis. (c) Similar configuration but with {-211}* domain walls (one of them is perpendicular to $y'$ axis). On the top right side, the respective orientation of the parent cubic perovskite unit cell is plotted along with the lattice vectors $a_c$, $b_c$ and $c_c$, while on the bottom right, the same structure is redrawn within the hexagonal unit cell of rhombohedral BaTiO$_3$ ($a_h = (0, 1, -1)$ and $b_h = (-1, 0, 1)$ form a 120-degree angle). The blue spheres stand for the Pb or Ba atoms, the green ones are the Ti atoms, and the red ones are the O atoms.

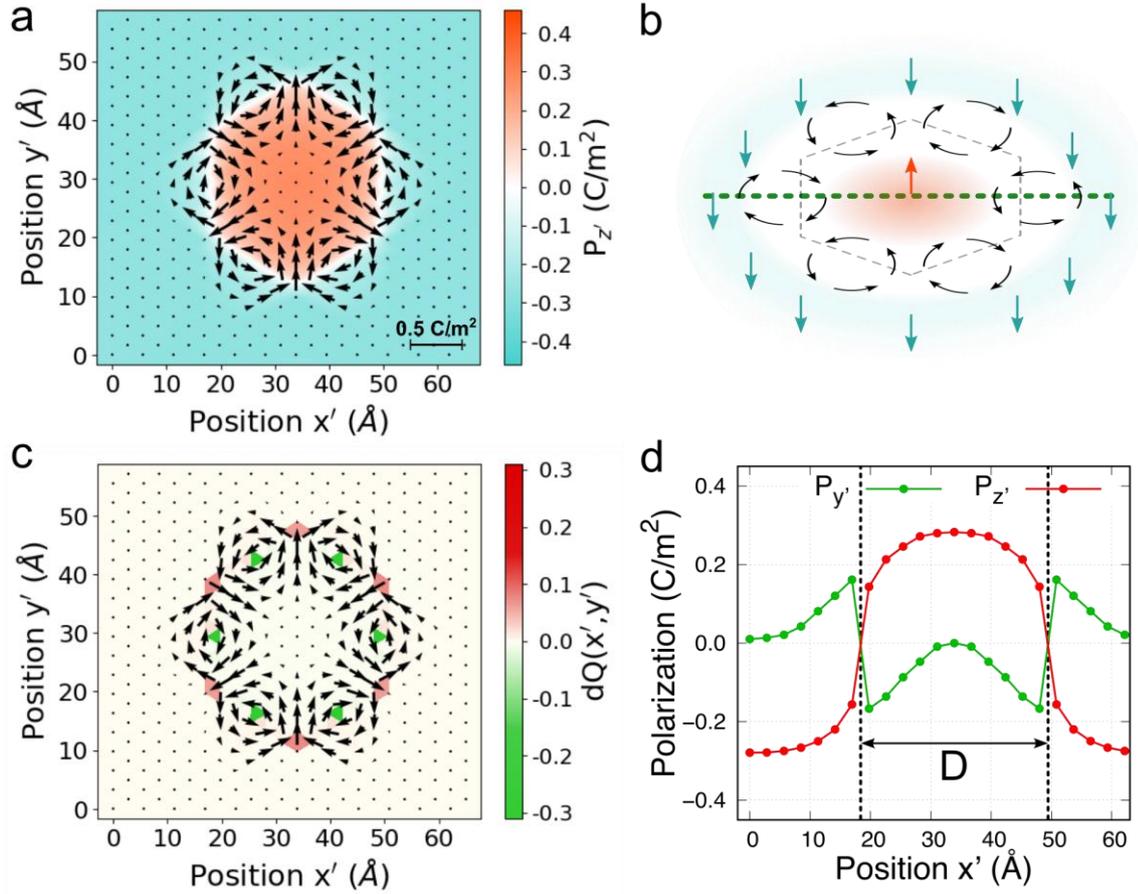

**Fig. 2. A relaxed antiskyrmion nanodomain in rhombohedral BaTiO$_3$.** (a) Section through the N = 11 antiskyrmion polarization configuration obtained by relaxation of a topologically trivial bubble nanodomain with the hexagonal section of Fig. 1b. Local polarization is calculated for each Ti site and projected along the *z'* axis. Non-negligible in-plane polarization components are marked by arrows, the out of plane components are indicated by color. (b) Simplified sketch of the same polarization pattern. The orange and turquoise arrows represent the nanodomain and the matrix. Curved arrows are indicating the six vortices of in-plane polarization part. (c) Topological charge density distribution *q(x', y')dS* superposed with the in-plane polarization. Values of *q(x', y')dS* are calculated separately for each individual triangular plaquette within the hexagonal sublattice of the projected Ti positions. (d) Transversal polarization components $P_{x'}$, $P_{z'}$ along the green dashed line of panel (b) with indicated skyrmion size *D*.

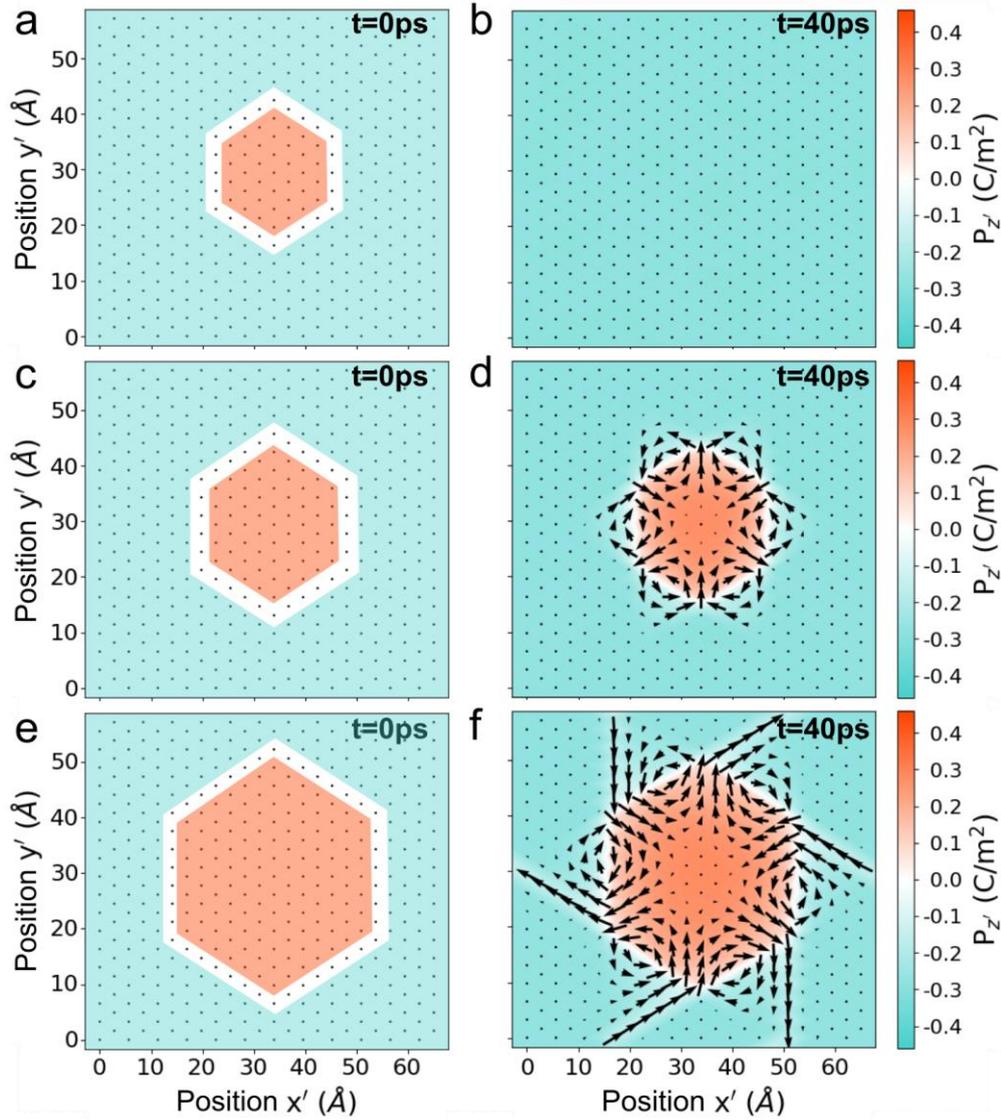

**Fig. 3. Formation of antiskyrmions with different sizes.** Left panels (a, c, e) show the initial nanodomain with a topologically trivial colinear polarization configuration. From top to bottom, the size parameter is $N = 7$, $N = 9$ and $N = 13$. Right panels (b, d, f) show the respective relaxed configuration after 40 ps of MD relaxation at $T = 1$ K.

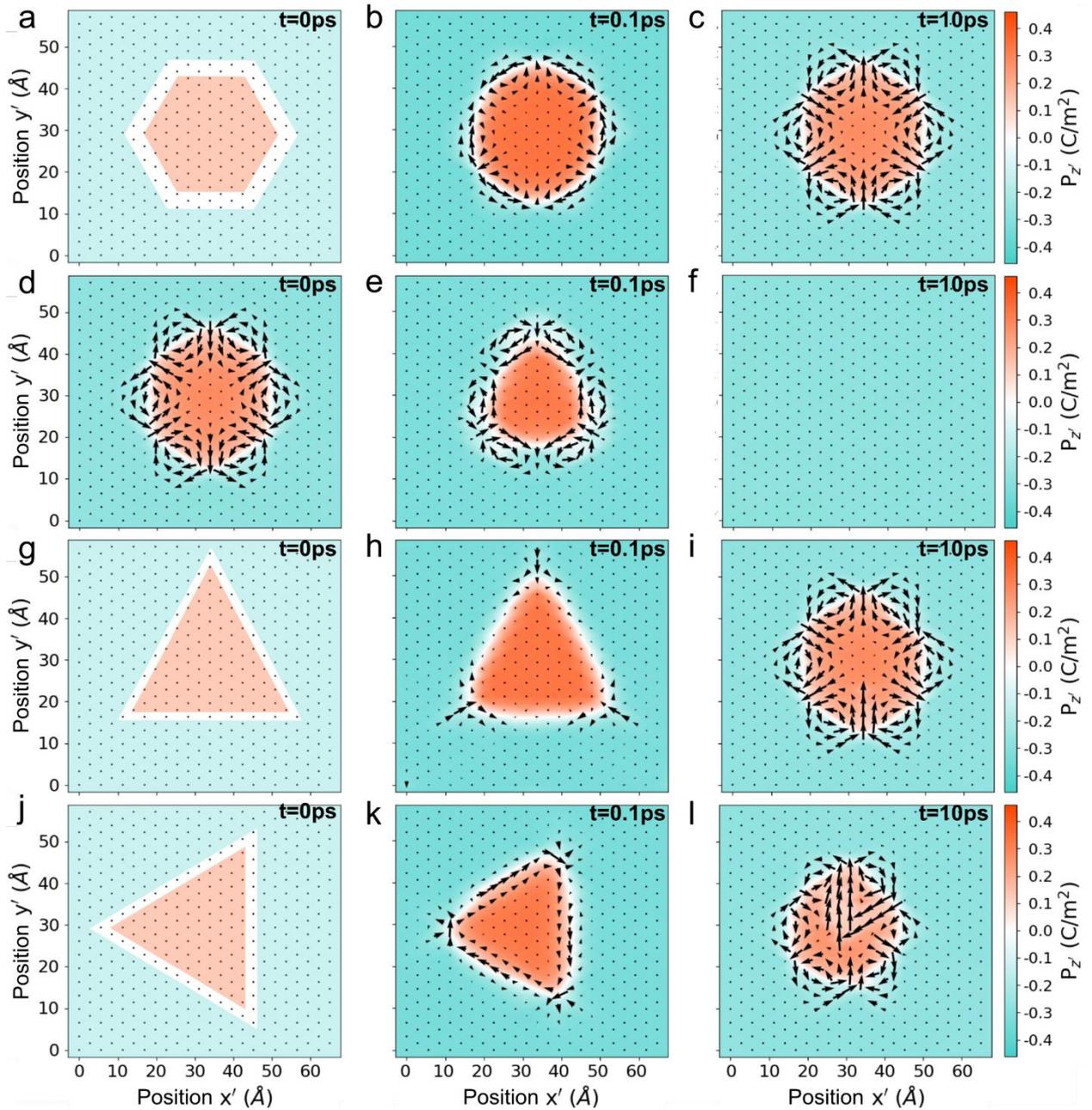

**Fig. 4. Decay of alternative initial configurations.** Left panels show the initial configurations, middle panels show the structures after 0.1 ps of MD evolution and the right panels show the close to final equilibrium configuration registered after 10 ps of MD evolution. (a, b, c) – configuration with R180{-211} domain walls transform to that with R180{01-1} domain walls. (d, e, f) – configuration similar to Fig. 2a but with inverted in-plane polarization decays into the single domain. Panel (g, h, i) and (j, k, l) show evolutions from colinear configurations with R180{-211} and R180{01-1} domain walls, respectively.